\documentclass[journal]{IEEEtran}
\usepackage{graphicx}
\usepackage{graphbox}
\usepackage{wrapfig}
\usepackage{gensymb}
\usepackage{amsmath}
\usepackage{cite}
\usepackage{url}
\usepackage{placeins}

\begin{document}

\title{Separation of two electromagnetic or electromagnetic - hadronic showers in CALICE SiW ECAL and ILD}

\author{Kostiantyn Shpak,~on behalf of the CALICE Collaboration

\thanks{K. Shpak is with the Laboratoire Leprince-Ringuet, Ecole polytechnique, Palaiseau cedex, 91128 France (e-mail: kostiantyn.shpak@llr.in2p3.fr).}%

\thanks{Talk presented at the International Workshop on Future Linear Colliders (LCWS2017), Strasbourg, France, 23-27 October 2017. C17-10-23.2.}
}

\maketitle

\thispagestyle{empty}

\begin{abstract}
CALICE collaboration is developing highly granular calorimeters suitable for individual reconstruction of particles in the jets using Particle Flow Algorithms. Such calorimeters should provide the best jet energy resolution at future high energy $e^{+}e^{-}$ colliders. At high jet energies, typically above 70-100~GeV, the jet particle showers start to overlap, and the resolution is determined by the ability to separate them. Here, we present the results on the separation of two overlapping electromagnetic or electromagnetic - hadronic showers obtained by mixing of the single shower events collected with CALICE SiW ECAL and AHCAL physics prototypes during beam tests at CERN'07 and FermiLab'11, and using International Large Detector (ILD) Monte Carlo simulations. We use three available PFA reconstruction programs (Pandora, Garlic and Arbor).
\end{abstract}


\section{Introduction}

Physics goals of the future high energy $e^{+}e^{-}$ colliders, such as ILC, CLIC, FCCee or CEPC, can be reachable with detectors optimized for the application of Particle Flow reconstruction Algorithms~(PFA)~\cite{Brient:2002gh}.  Initially, the idea of the PFA was proposed purely on the basis of Monte-Carlo~(MC) simulations, and up to now it was never verified with jets in the operational physics experiment. There are some attempts to apply PFA to large LHC experiments, but their reconstruction environment and detectors are not optimized for PFA. Highly granular calorimeters suitable for the application of the PFA are being developed within the CALICE collaboration~\cite{CALICEcollaboration}.

Designed detectors for future $e^{+}e^{-}$ collider, like International Large Detector~(ILD)~\cite{Behnke:2013lya}, should provide 3-4\% of Jet Energy Resolution~(JER) for 45-250~GeV jets to distinguish W and Z bosons decaying hadronically. JER of the low energy jets is limited by the intrinsic stochastic resolution of the hadronic calorimeter and scales with energy as $1/\sqrt{E}$. However, for higher jet energies~(above $E_{jet}>100$~GeV) performance of the PFA is limited by confusion~\cite{Marshall:2013bda}, impossibility to correctly associate calorimetric hits with corresponding jet particles, causing degradation of JER.

Another important requirement placed on the ILC detectors is possibility to distinguish hadronic $\tau$-decay modes~\cite{Tran:2015nxa}: $\tau^{+} \rightarrow \pi^{+} \bar{\nu}_{\tau}$, $\rho^{+}(\pi^{+}\pi^{0}) \bar{\nu}_{\tau}$ and $a_{1}^{+}(\pi^{+}\pi^{0}\pi^{0})  \bar{\nu}_{\tau}$, by distinguishing $\pi^{0}\rightarrow \gamma \gamma$ decays. High energy $\pi^{0}$ are characterized by the small opening angle between decay photons and, correspondingly, by small distances between photons on the ECAL front face.

\begin{figure}[t]
	\centering
 	\includegraphics[width=.35\textwidth]{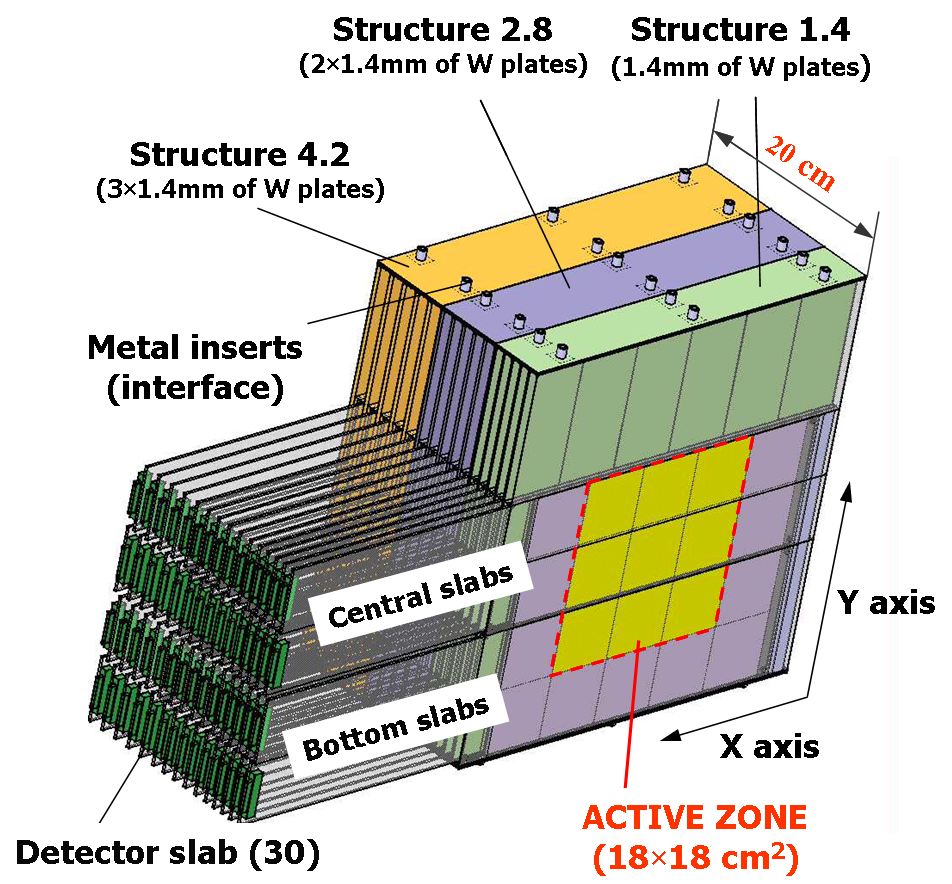}
 	\caption{Schematic 3D view of CALICE SiW ECAL physics prototype.} 
 	\label {calice}
\end{figure}

\begin{figure}[t]
	\centering
 	\includegraphics[width=.2\textwidth]{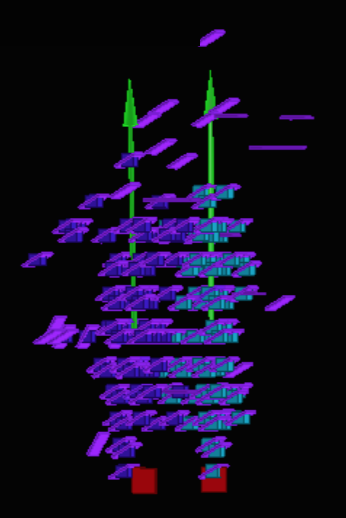}
 	\caption{One example of the correct reconstruction of two overlapping electromagnetic showers initiated by 4~GeV positrons separated by 2~cm using Garlic program.} 
 	\label {2particle}
\end{figure}

In order to minimize the confusion and improve the intrinsic detector resolution, PFA requires calorimeters (both ECAL and HCAL) with high segmentation in both transverse and longitudinal directions. Probably, the most suitable ECAL technology for PFA is a silicon-tungsten sampling calorimeter. Tungsten~(W) is an optimal choice for the absorber as it has relatively small Moliere radius ($R_{M}^{W}\approx9.3$~mm) ensuring the small transverse size of the electromagnetic~(EM) showers. For W the large ratio between nuclear interaction length ($\lambda_{int}^{W} \approx 99.5$~mm) and radiation length ($X_{0}^{W} \approx 3.5$~mm), $\lambda_{int}^{W} / X_{0}^{W} \approx 28.4$, cause later starts of hadronic showers in comparison with electromagnetic showers, ensuring good separation between EM and hadronic showers in the longitudinal direction. Silicon as active layer readout material is expensive but offers numerous advantages. It is easily segmentable so that its granularity is limited only by the number of channels in the front-end electronics. It is reliable, stable in time, not sensitive to any environmental changes (e.g., to the temperature) and can work in the strong magnetic field. Silicon detector should be easy to calibrate on the cell-by-cell level at each stage of detector operation. Active layer transverse granularity close to $\sim 5 \times 5$~mm$^2$ is sufficient to reach physics goals of the ILC. With the SiW technology, one can achieve the lowest level of ECAL systematic errors and obtain an excellent granularity for PFA.

Pattern recognition capabilities of highly granular calorimeters and PFA have been studied mainly only using MC simulations. CALICE TB data have been used for confirmation of high physics potential of the PFA calorimetry. In this study, confusion is studied in the reconstruction of two overlapped showers obtained by event mixing. Only cases relevant to SiW~ECAL performance are considered:
\begin{itemize}
	\item separation of two electromagnetic showers (EM-EM), relevant for $\pi^{0}$ reconstruction;
	\item separation of hadronic and electromagnetic showers (hadron-EM), relevant for JER performance of the detector.
\end{itemize}

\section{Detectors and software}

Study of EM-EM (hadron-EM) separation is based on the data collected with CALICE physics prototype of SiW~ECAL (SiW~ECAL + AHCAL) during beam test campaign at FermiLab'11~(CERN'07). Complete SiW~ECAL prototype, see Fig.~\ref{calice}, has sandwiched structure of 30~active~Si and 30~W~absorber layers. Transversely each silicon layer has an active area of $18 \times 18$~cm$^2$ segmented into $3 \times 3$ wafers with $1 \times 1$~cm$^2$ cells, leading to 9720 channels in a full prototype. Absorber thickness varies across the calorimeter, the first 10 W layers are $1.4$~mm thick (corresponding to $0.4X_{0}$), next 10 in the middle are twice thicker, $2.8$~mm ($0.8X_{0}$), and the last 10 layers have a thickness of $4.2$~mm ($1.2X_{0}$). The total depth of the SiW~ECAL prototype corresponds to $24X_{0}$ with an overall ECAL thickness of 20~cm. AHCAL in CERN'07 tests is composed of 38 layers of the highly segmented scintillator and 38 steel absorber planes. AHCAL has a granularity of $3 \times 3$~cm$^2$ in the central part of 30 front-most layers, while rear and peripheral regions of the HCAL have worse granularity - $6 \times 6$~cm$^2$ or $12 \times 12$~cm$^2$. A detailed description of the CALICE SiW~ECAL and AHCAL prototypes and some physics results can be found in~\cite{Anduze:2008hq, collaboration:2010hb, Adloff:2011ha}.

We compare results obtained with TB data and MC simulations (QGSP\_BERT physics list) in corresponding CALICE geometries. In addition, separation studies are also performed in the full ILD detector models with standard $5 \times 5$~mm$^2$ or finer $2.5 \times 2.5$~mm$^2$ ECAL cells and with AHCAL ($30 \times 30$~mm$^2$ cells with analog readout, when the signal is proportional to the energy deposit in each calorimeter cell) or with SDHCAL ($10 \times 10$~mm$^2$ cells with semi-digital 2-bit readout, it counts hits above three different thresholds per cell). 

Three available PFA reconstruction programs are used for both CALICE and ILD events:
\begin{itemize}
	\item Pandora~\cite{Marshall:2013bda} v00-14 and v02-04, standard ILD reconstruction tool providing the best JER, optimized for ILD with AHCAL;
	\item Garlic~\cite{Jeans:2012jj} v2.11 and v3.0.3, designed to reconstruct photons, uses only ECAL information;
	\item Arbor~\cite{Ruan:2014paa} March'15 version, an alternative to Pandora full event reconstruction algorithm based on treelike clustering, best for ILD with SDHCAL.
\end{itemize}
The obtained results are new, full details of this study can be found in CALICE Analysis Note CAN-057~\cite{shpak:CAN057}. Confusion limits of the PFA algorithms are tested in a more direct way than before with all shower details taken from the data. The best performance of the PFA tools is demonstrated with MC simulations in the standard ILD geometry.

\section{Method}

Used PFA reconstruction programs are developed for the reconstruction of the full events recorded in the ILD detector. These programs are best optimized for ECAL with $5 \times 5$~mm$^2$ cells. To apply available PFA tools for CALICE prototypes data, the laters are first converted to the ILD geometry by turn and radial shift.

The principal difference between CALICE setups and ILD is the absence of tracking device and magnetic field in TB. This affects shapes of the recorded showers, but the influence of this effect on the final results is not studied. Since it is difficult to get monochromatic high energy $\gamma$ beam, $\gamma$ showers can be emulated with $e^{+}$ collected during TB. $\pi^{+}$ showers are complemented with tracks after placing into ILD geometry.

Details of the CALICE ECAL and AHCAL calibration procedure can be found in~\cite{CAN001, collaboration:2010hb, shpak:CAN057}. The energy of the TB events is estimated as:
\begin{multline}
	E=\sum_{j=0}^{j=9}E^{ECAL}_{j} + 2 \sum_{j=10}^{j=19}E^{ECAL}_{j} + 3 \sum_{j=20}^{j=29}E^{ECAL}_{j} \\ 
	+ \sum_{k=0}^{j=37}E^{AHCAL}_{k} \  ,
\end{multline}
where $E^{ECAL}_{j}$ is energy recorded in $j$th layer of ECAL and $E^{AHCAL}_{k}$ - in $k$th layer of AHCAL (last AHCAL term exists only for $\pi^{+}$). Coefficients 1, 2 and 3 placed in front of ECAL-related terms correspond to single, double and triple tungsten thicknesses in front of corresponding silicon layers. 

Events in the used $e^{+}$ and $\pi^{+}$ beam test samples passed additional selection criteria to suppress events with following effects:
\begin{itemize}
    \item events with sensors affected by so-called plane and "square"-events~\cite{Shpak:2016anm};
    \item multiple particles, only events where the number of hits (energy) in the first ECAL layer is exactly one (below 2 MIPs) are selected;
    \item showers with significant transverse energy leakage, events centered in the central $4 \times 4$~cm$^2$ part of the central ECAL wafer pass this criterion;
    \item $\mu^{+}$, $\pi^{+}$, $p$ ($\mu^{+}$ and $\pi^{+}$ with late showers) in $e^{+}$ ($\pi^{+}$) samples are suppressed by a cut on the total energy in ECAL (ECAL+HCAL).
\end{itemize}

MC generated samples of $e^{+}$ and $\pi^{+}$ do not have issues related to the experimental environment, but same cuts as for TB events are applied. Separation studies performed with $e^{+}$ emulating $\gamma$ are compared to real $\gamma$ simulations, the selection procedure for $\gamma$ is almost identical to $e^{+}$ with a modified cut on the number of hits (energy) in the first ECAL layer: 0 or 1 (below 4~MIPs). Detailed description of the applied selection criteria and corresponding distributions can be found in~\cite{shpak:CAN057}.

In case of ILD, $\gamma$ and $\pi^{+}$ samples with energies similar to CALICE TB energies are simulated, no cuts applied. $\pi^{+}$ for the Pandora (Arbor) analysis are simulated in the ILD geometry with AHCAL (SDHCAL). Garlic does not use HCAL information, but to clarify $\pi^{+}$ events are the same as for Pandora analysis.

Two overlapping showers are obtained by event mixing. The only normal incidence of particles has been studied. Before the mixing one EM shower is shifted by 1, 2, 3,... cells (only for CALICE, as ILD samples are broad enough to avoid this procedure).

The ECAL entering point is calculated as shower barycenter for EM showers ($e^{+}$, $\gamma$). For hadrons, the entering point is defined differently, as hit position in the first ECAL layer for CALICE, and, for ILD, the entering point can be measured with better precision with the tracker.

EM-EM or hadron-EM events are selected for the analysis if initial EM showers before mixing are reconstructed as single $\gamma$ showers. In addition, it is required to reconstruct initial $\pi^{+}$ showers (hadron-EM case) used for Pandora and Garlic analysis as single $\pi^{+}$ without any neutrals with Pandora. For Arbor the selection requirement on the initial $\pi^{+}$ is softer, it should be reconstructed as $\pi^{+}$ but unphysical neutrals in HCAL are allowed.

Reconstruction of mixed EM-EM (hadron-EM) event is considered as successful if exactly two (one) $\gamma$ are reconstructed and their energies and transverse shower barycenter coordinates are within $\pm 20\%$ and $\pm 5$~mm of the corresponding values of the initial reconstructed $\gamma$ before mixing. Also for Pandora (Arbor), it is required to reconstruct $\pi^{+}$ in addition to $\gamma$ in the $\pi^{+}$-EM events after mixing.

The probability of the correct reconstruction of two showers $P(\Delta)$ is determined as a function of the distance between the particle entering points after the mixing,~$\Delta$. Since the size of the beam was comparable to the granularity of the CALICE setup ($1 \times 1$~cm$^2$ and $3 \times 3$~cm$^2$ for the SiW~ECAL and central AHCAL parts, respectively), all $\Delta$ separation distances may be studied in this way, and $P(\Delta)$ is obtained as a continuous curve.

An example of the correct reconstruction of two overlapping electromagnetic showers by Garlic program is shown in Fig.~\ref{2particle}.

\begin{figure*}[!htb]
	\centering
  	\includegraphics[width=.99\textwidth]{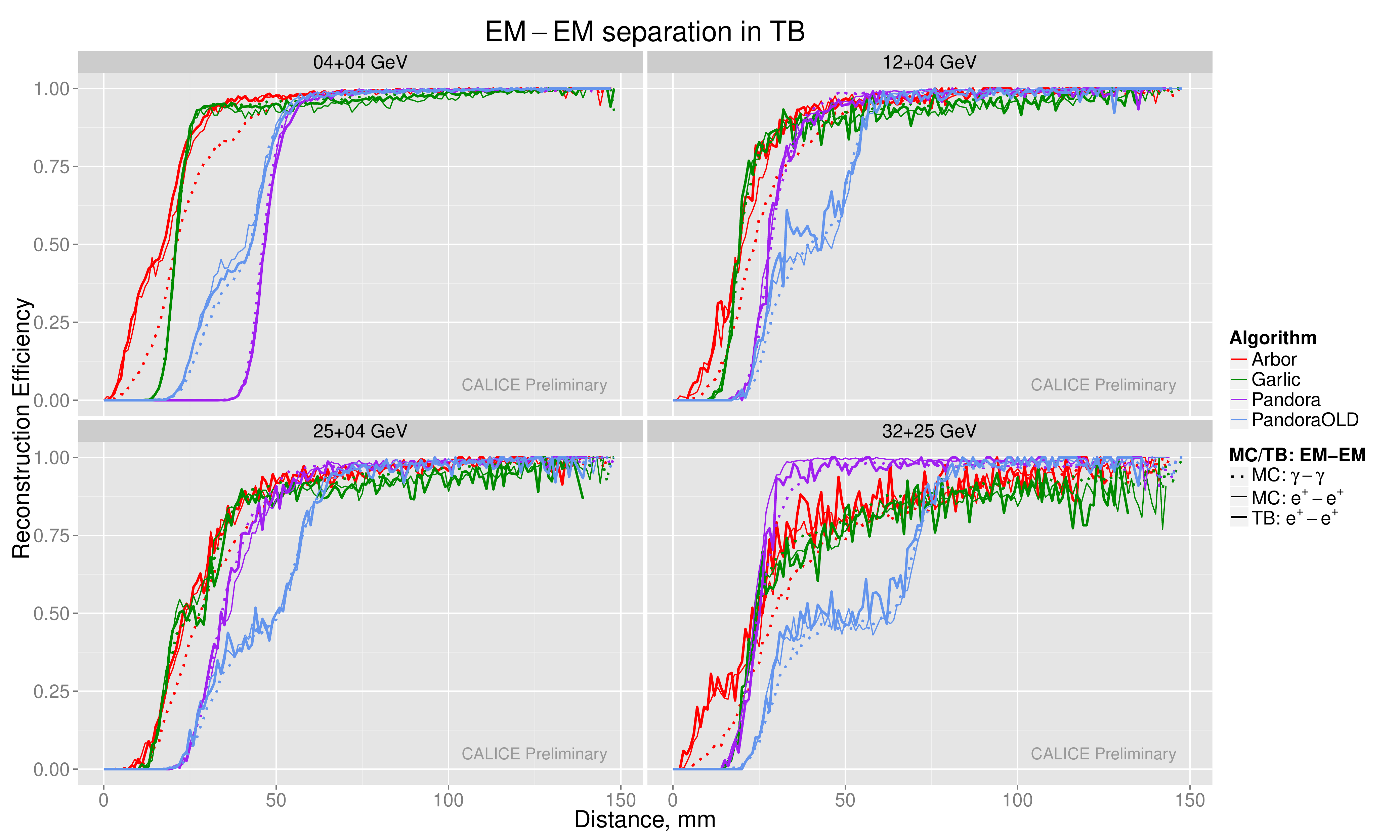}
    \caption{The probability of correct reconstruction of two overlaid EM showers ($e^{+} - e^{+}$ or $\gamma - \gamma$; 4+4, 12+4, 25+4 and 32+25~GeV energy pairs) as a function of the distance between them for CALICE SiW ECAL with $1 \times 1$~cm$^2$ cells.} 
	\label {separationphotonTB}
\end{figure*}
\begin{figure*}[!htb]
	\centering
	\includegraphics[width=.99\textwidth]{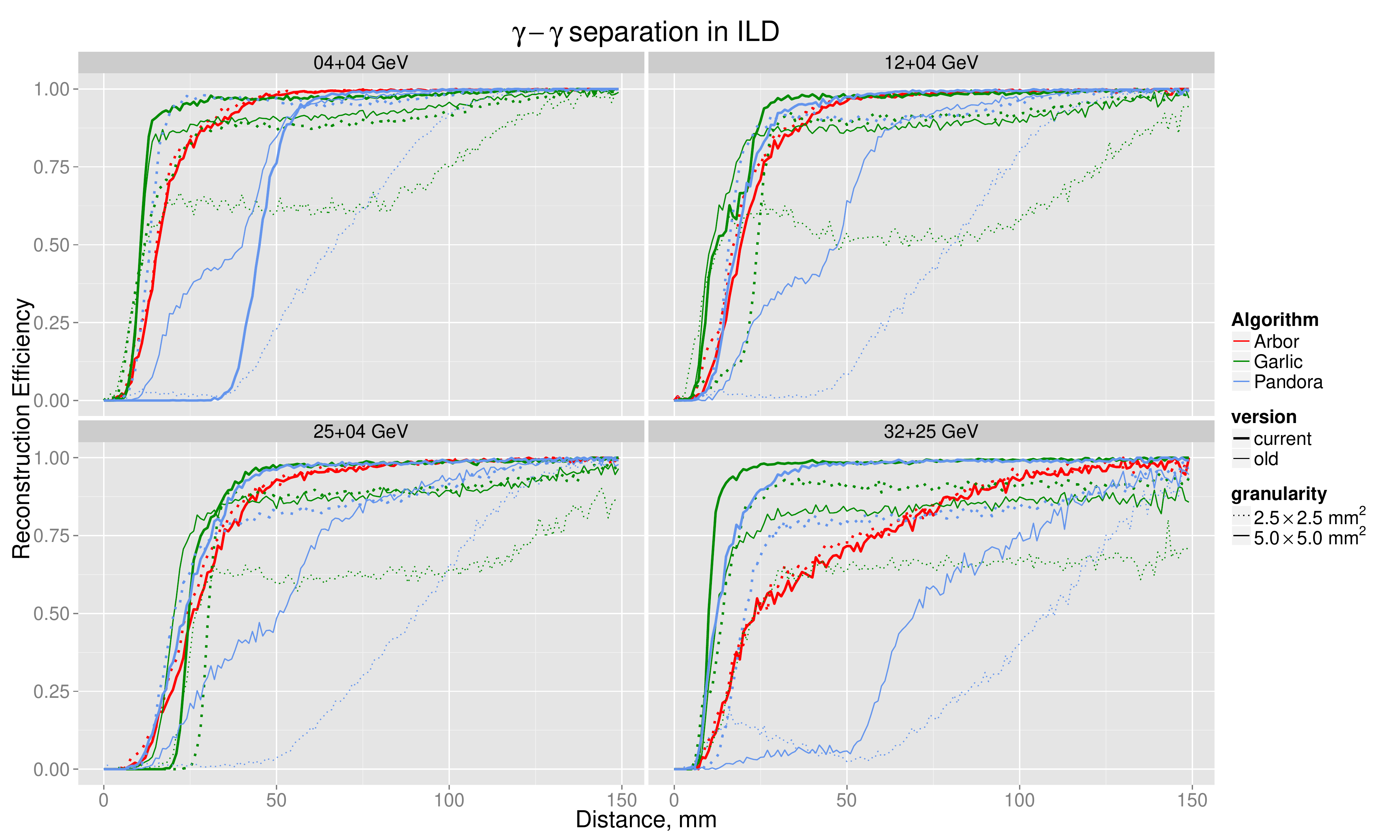}
	\caption{The probability of correct reconstruction of two overlaid $\gamma$ showers (4+4, 12+4, 25+4 and 32+25~GeV energy pairs) as a function of the distance between them for ILD with standard $5 \times 5$~mm$^2$ and finer $2.5 \times 2.5$~mm$^2$ ECAL cells.} 
	\label {separationphotonILD}
\end{figure*}
\begin{figure*}[!htb]
	\centering
  	\includegraphics[width=.99\textwidth]{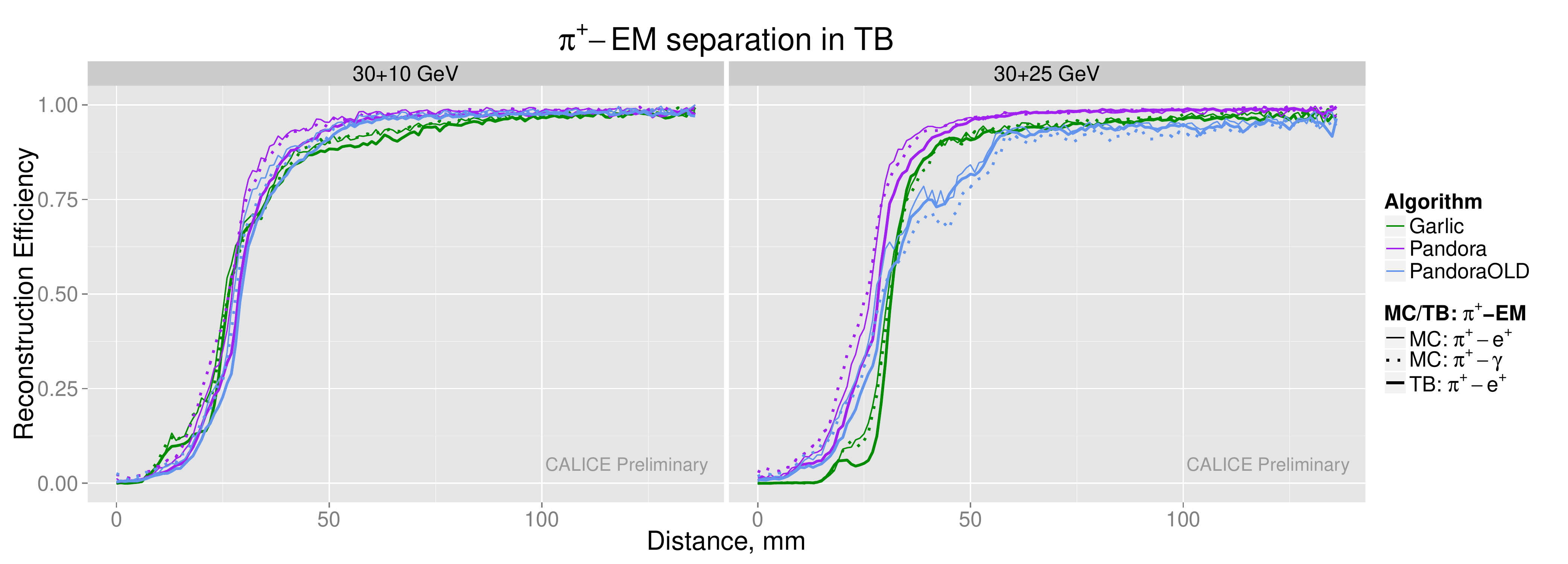}
    \caption{The probability of correct reconstruction of overlaid hadronic and EM showers ($\pi^{+} - e^{+}$ or $\pi^{+} - \gamma$; 30+10, 30+25 GeV) as a function of the distance between them for CALICE setup.} 
	\label {separationhadronsTB}
\end{figure*}
\begin{figure*}[!htb]
	\centering
	\includegraphics[width=1.02\textwidth]{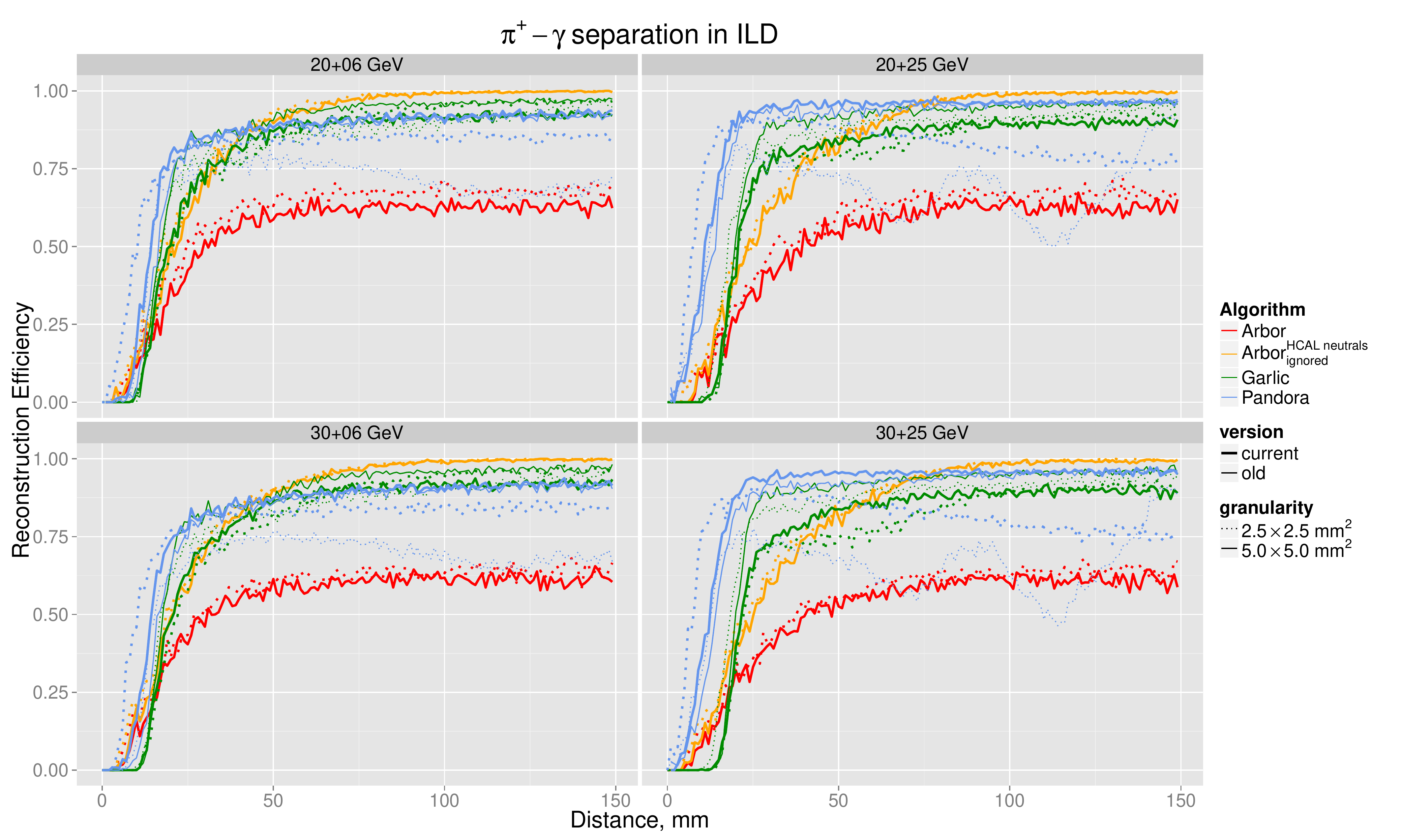}
 	\caption{The probability of correct reconstruction of overlaid $\pi^{+}$ and $\gamma$ showers (20+6, 20+25, 30+6 and 30+25~GeV) as a function of distance between them for ILD.} 
 	\label {separationhadronsILD}
\end{figure*}

\section{Results}
\subsection{Reconstruction efficiency}

Obtained probabilities $P(\Delta)$ for all studied energy pairs are presented in Figs.~\ref{separationphotonTB},~\ref{separationphotonILD} for the EM-EM case and in Figs.~\ref{separationhadronsTB},~\ref{separationhadronsILD} for hadron-EM separation. The EM-EM (hadron-EM) separation is studied using only ECAL (both ECAL and HCAL) information. Garlic does not provide full event reconstruction, as it does not use HCAL hits for the analysis. Garlic v2.11, the only version of Garlic used for CALICE data, is particularly tuned for the TB data analysis. It does not check the consistency of cluster direction with direction to IP, also for this version the veto region around extrapolated tracks is enlarged to 15~mm (10~mm is default). Since Arbor is optimized for highly granular HCAL ($1 \times 1$~cm$^2$ cells), the CALICE Fig.~\ref{separationhadronsTB} does not contain the Arbor curve. As one can see from the CALICE related Fig.~\ref{separationphotonTB}~(\ref{separationhadronsTB}), there is good agreement between TB and MC curves in case of $e^{+}-e^{+}$ ($\pi^{+}-e^{+}$) separation. Also, Garlic and Pandora do not show any significant difference between $e^{+}-e^{+}$ ($\pi^{+}-e^{+}$) and $\gamma - \gamma$ ($\pi^{+} - \gamma$) separation studied with MC, it means that emulation of $\gamma$ with $e^{+}$ do not cause any significant biases. However, this is not the case for Arbor, and one can see that performance of $e^{+}-e^{+}$ curve is slightly better than the $\gamma - \gamma$. This difference is caused by our selection procedure, for each selected $e^{+}$ it is mandatory to have a single hit in the first ECAL layer, while $\gamma$ can start interaction later, this single hit serves as a perfect seed for Arbor treelike clustering. It means that $\gamma - \gamma$ curve should be considered as the real performance of the Arbor.

The best performance of the available PFA tools is demonstrated in the standard ILD geometry with $5 \times 5$~mm$^2$ ECAL~(see solid curves in Figs.~\ref{separationphotonILD} and~\ref{separationhadronsILD}). Garlic provides the best performance in $\gamma - \gamma$ separation among all used algorithms. One can see significant improvement in the Pandora and Garlic performances in comparison with their older versions in case of $\gamma - \gamma$ separation. Pandora can be additionally tuned in case of low energy EM-EM pair (4+4~GeV) as it is demonstrated by Arbor and Garlic. Arbor performance is competitive with Garlic, but it tends to over-split showers in case of high photon energies (32+25~GeV pair).

Pandora is the best program for $\pi^{+}-\gamma$ separation in the ILD geometry with the standard $5 \times 5$~mm$^2$ ECAL granularity~(see solid curves in Fig.~\ref{separationhadronsILD}). The performance of the older Garlic v2.11 is slightly better than the new v3.0.3, as the new version is particularly tuned for the $\gamma$ finding, so it finds more $\gamma$ candidates, and its $\pi^{+}-\gamma$ separation performance is slightly degraded. Arbor tends to over-split $\pi^{+}$ showers in HCAL, that is the reason why the efficiency of $\pi^{+}-\gamma$ separation with Arbor is limited by ${\sim 70\%}$. If unphysical neutrals in the HCAL will be neglected, the performance of the Arbor can be shown with the orange curve in Fig.~\ref{separationhadronsILD}.

Performances of the Pandora and Garlic for both $\gamma-\gamma$ and $\pi^{+}-\gamma$ separation are significantly degraded for the ECAL with finer $2.5 \times 2.5$~mm$^2$ cells~(dotted curves in Figs.~\ref{separationphotonILD} and~\ref{separationhadronsILD}), while it is natural to expect some improvement of the performance. Nevertheless, it is seen that for some $\gamma - \gamma$ and $\pi^{+} - \gamma$ pairs for small distances below 10~mm the separation performance of Pandora is better for finer $2.5 \times 2.5$~mm$^2$ ECAL cells. It means that the settings of Pandora and Garlic are not optimal for $2.5 \times 2.5$~mm$^2$ ECAL granularity and additional tuning is needed. However, this is not the case for Arbor, and its performance is even slightly improved for the finer ECAL granularity.

\subsection{Classification of inefficiencies}

\begin{figure*}[!htb]
	\centering
	\begin{minipage}{0.93\linewidth}
 	\includegraphics[align=c, width=\textwidth]{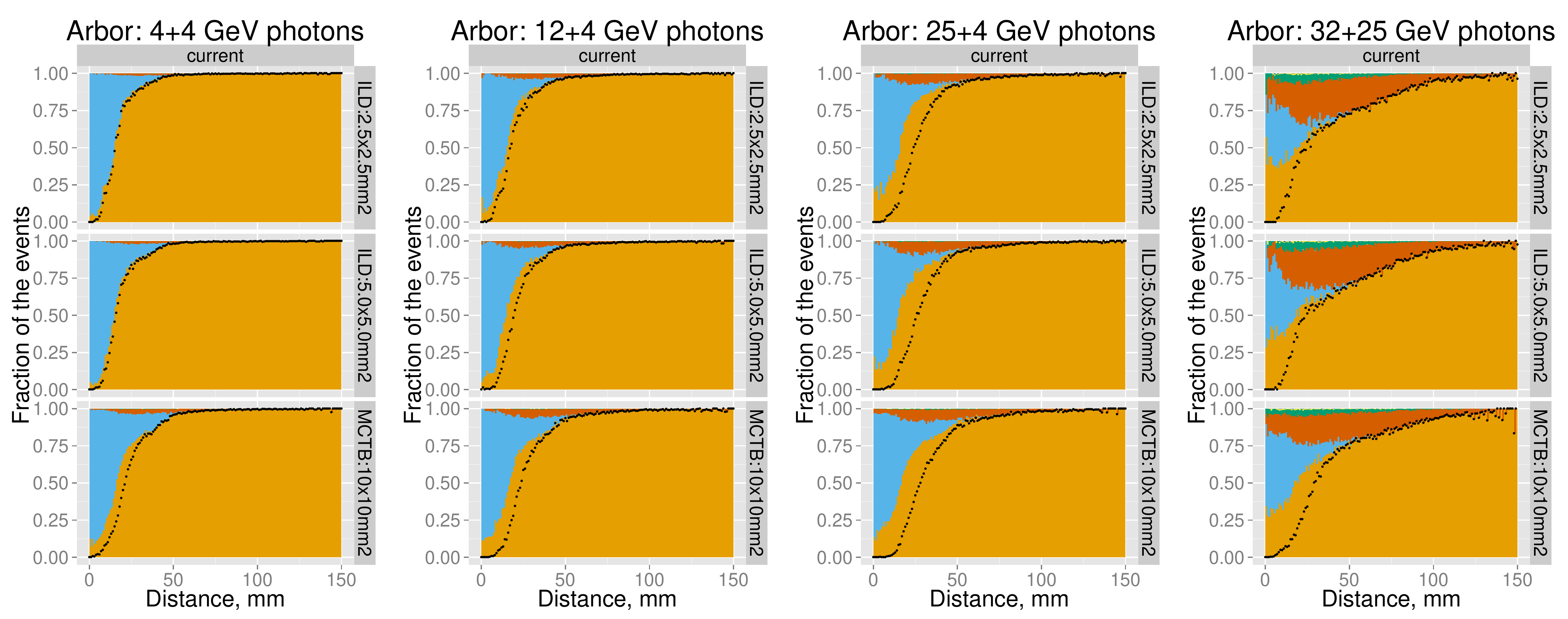}
 	\end{minipage}
 	\begin{minipage}{0.06\linewidth}
 	\includegraphics[align=c, width=\textwidth]{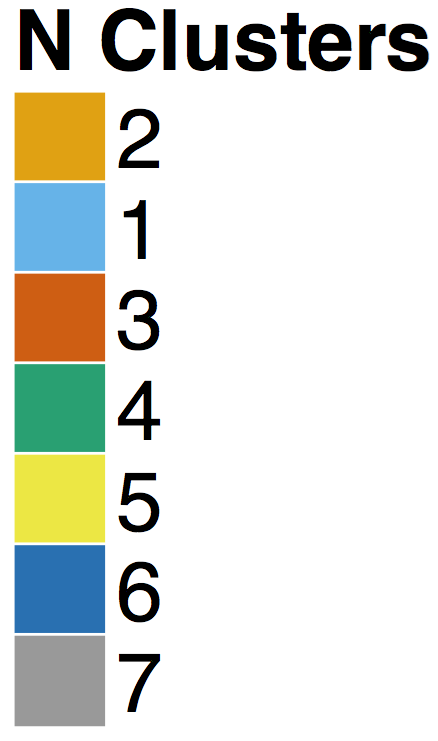}
 	\end{minipage}
 	\caption{Arbor reconstruction of 4+4, 12+4, 25+4 and 32+25~GeV $\gamma - \gamma$ events: the colored bands show the fraction of events where 1, 2, 3,... neutral clusters are reconstructed, regardless of their energies and positions. The black points show the efficiency of reconstruction of exactly two clusters with the energies and the positions within $\pm 20\%$ and $\pm 5~mm$, respectively, from their values in single $\gamma$ shower events, i.e. as in Figs.~\ref{separationphotonTB} and~\ref{separationphotonILD}.}
 	\label {2photonsArbor}
\end{figure*}

\begin{figure*}[!htb]
	\centering
	\begin{minipage}{0.88\linewidth}
 	\includegraphics[align=c, width=\textwidth]{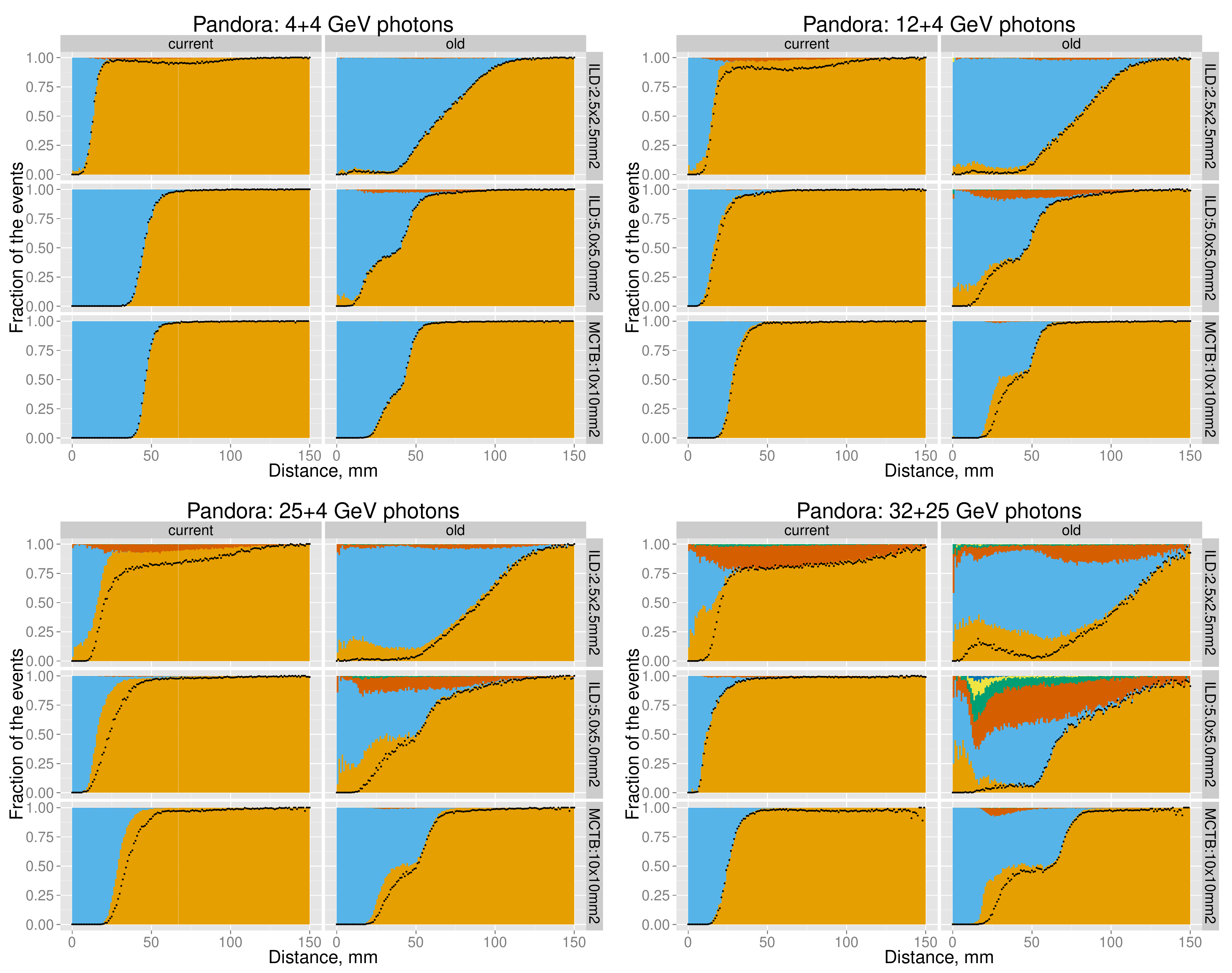}
 	\end{minipage}
 	\begin{minipage}{0.06\linewidth}
 	\includegraphics[align=c, width=\textwidth]{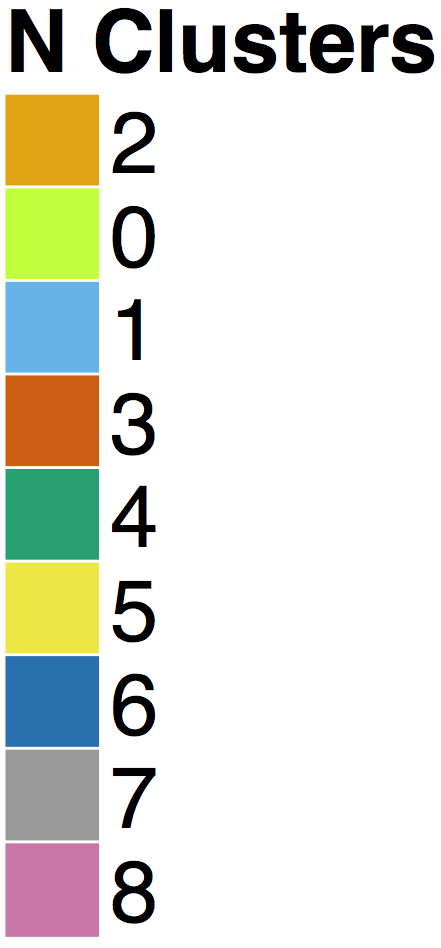}
 	\end{minipage}
 	\caption{Pandora reconstruction of 4+4, 12+4, 25+4 and 32+25~GeV $\gamma - \gamma$ events: the colored bands show the fraction of events where 1, 2, 3,... neutral clusters are reconstructed, regardless of their energies and positions.}
 	\label {2photonsPandora}
\end{figure*}

\begin{figure*}[!htb]
	\centering
	\begin{minipage}{0.88\linewidth}
 	\includegraphics[align=c, width=\textwidth]{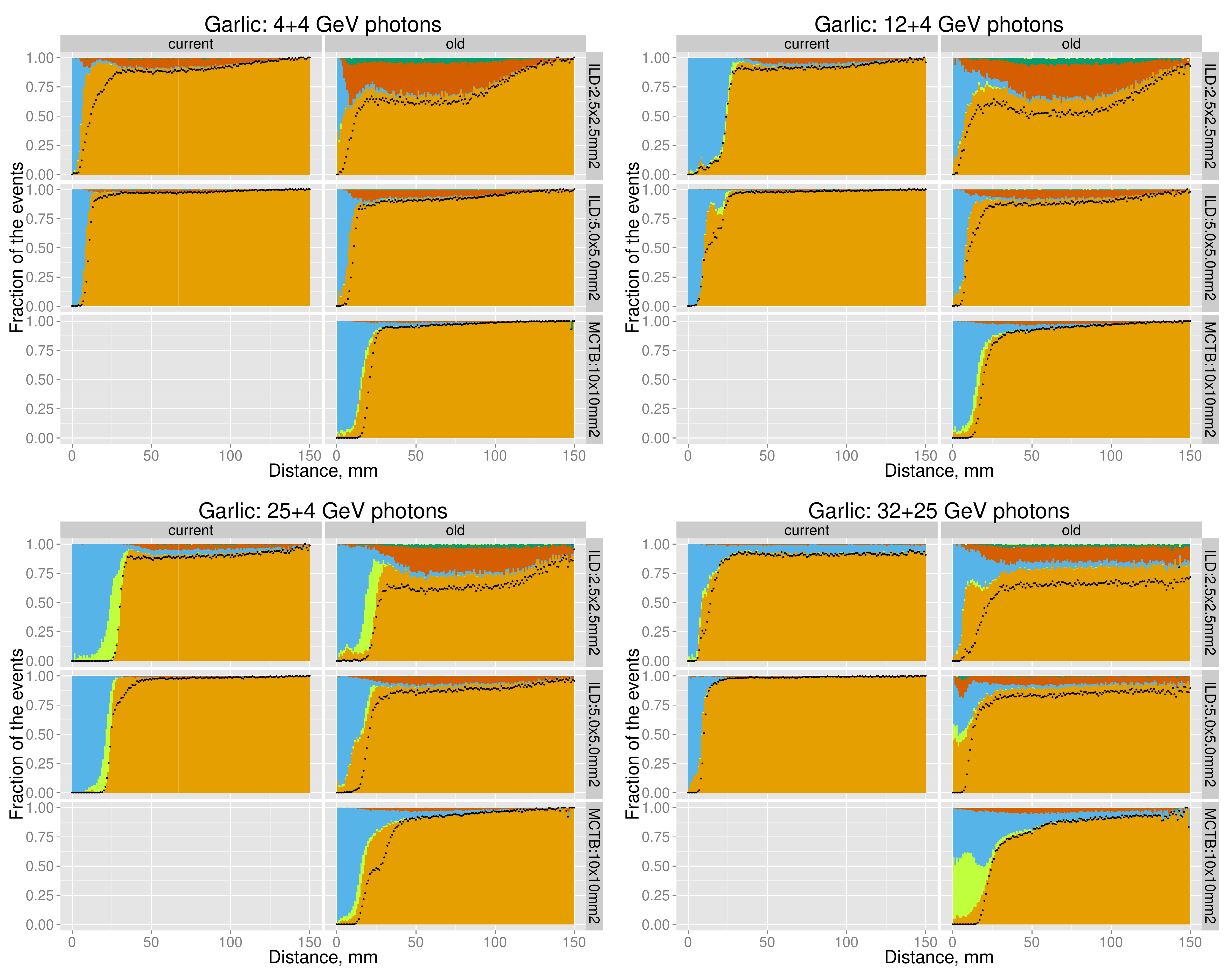}
 	\end{minipage}
 	\begin{minipage}{0.06\linewidth}
 	\includegraphics[align=c, width=\textwidth]{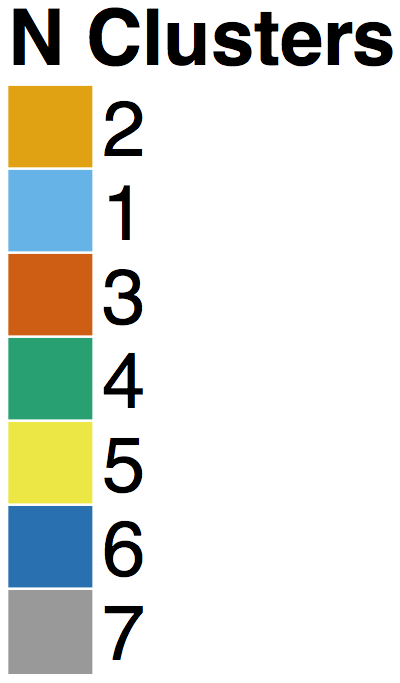}
 	\end{minipage}
 	\caption{Garlic reconstruction of 4+4, 12+4, 25+4 and 32+25~GeV $\gamma - \gamma$ events: the colored bands show the fraction of events where 1, 2, 3,... $\gamma$ are reconstructed, regardless of their energies and positions.}
 	\label {2photonsGarlic}
\end{figure*}

\begin{figure*}[!htb]
	\centering
	\begin{minipage}{0.88\linewidth}
 	\includegraphics[align=c, width=\textwidth]{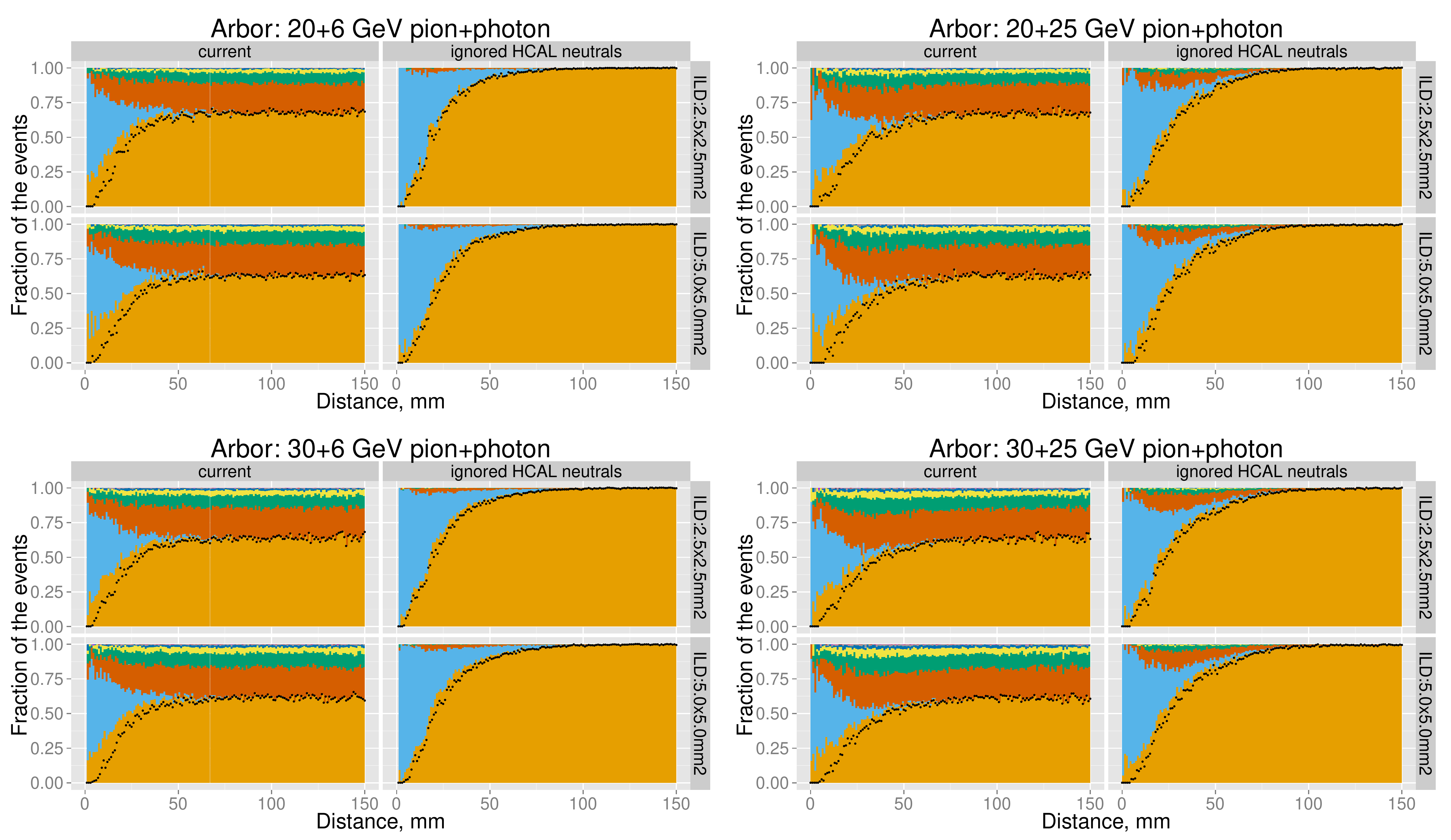}
 	\end{minipage}
 	\begin{minipage}{0.06\linewidth}
 	\includegraphics[align=c, width=\textwidth]{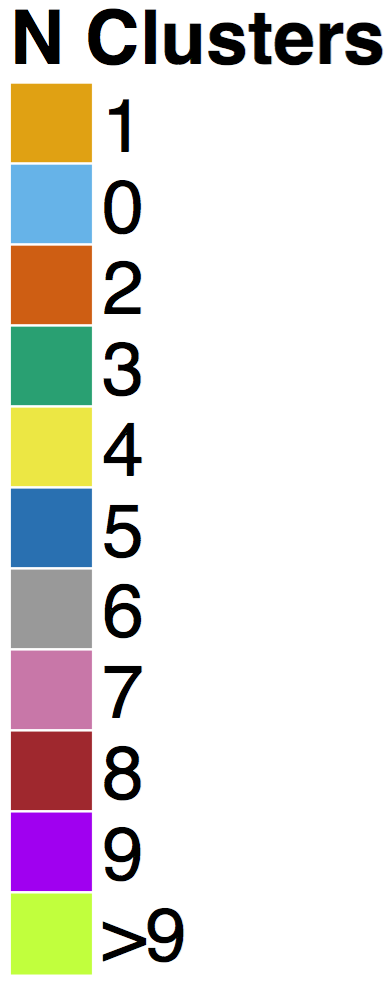}
 	\end{minipage}
 	\caption{Arbor~(SDHCAL) reconstruction of 20+6, 20+25, 30+6 and 30+25~GeV $\pi^{+}-\gamma$ events: the colored bands show the fraction of events with 0, 1, 2,... reconstructed neutral clusters, regardless of their energies and positions. The black points show the same efficiency as in Fig.~\ref{separationhadronsILD}. In the right column plots ("ignored HCAL neutrals") we ignore all HCAL neutral clusters, so that all inefficiencies are related to ECAL.}
 	\label {pionphotonArbor}
\end{figure*}

\begin{figure*}[!htb]
	\centering
	\begin{minipage}{0.88\linewidth}
 	\includegraphics[width=\textwidth]{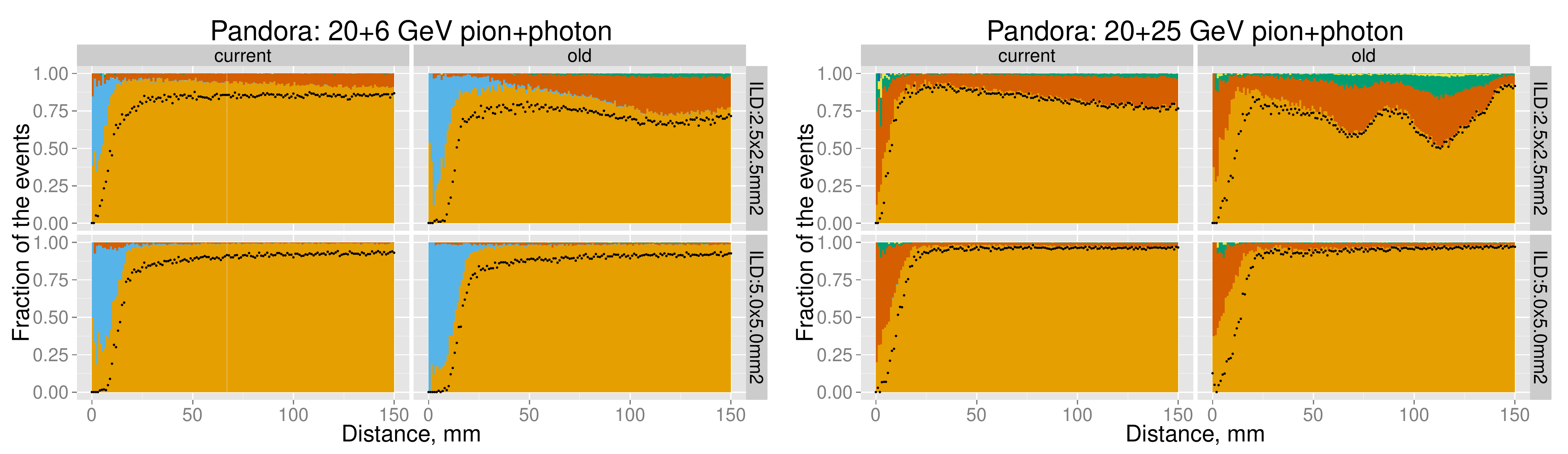}
 	\includegraphics[width=\textwidth]{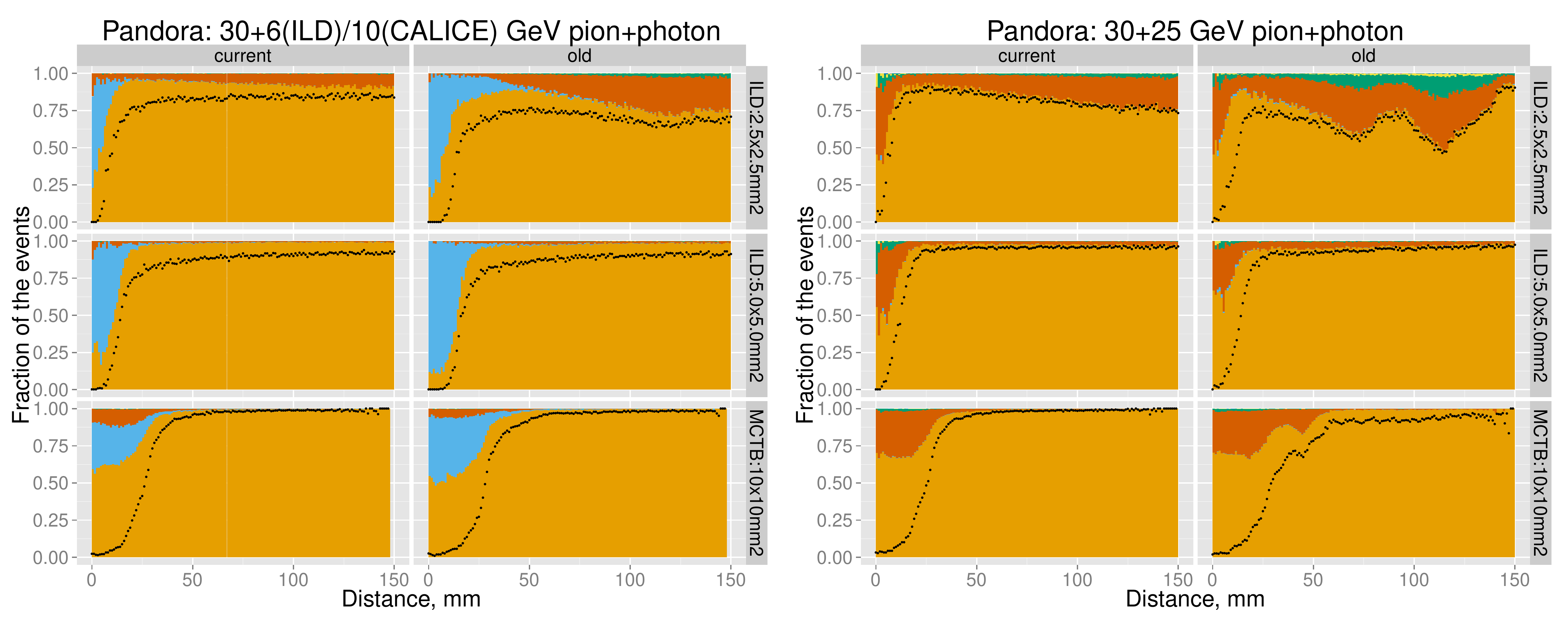}
 	\end{minipage}
 	\begin{minipage}{0.06\linewidth}
 	\includegraphics[align=c, width=\textwidth]{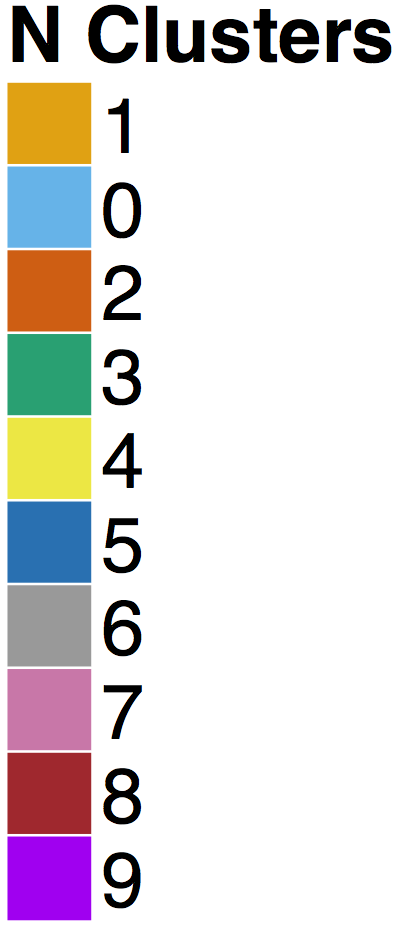}
 	\end{minipage}
 	\caption{Pandora~(AHCAL) reconstruction of 20+6, 20+25, 30+6~(10), 30+25 GeV $\pi^{+}-\gamma$ events: the colored bands show the fraction of events with 0, 1, 2, ... reconstructed neutral clusters, regardless of their energies and positions.}
 	\label {pionphotonPandora}
\end{figure*}

\begin{figure*}[!htb]
	\centering
	\begin{minipage}{0.88\linewidth}
 	\includegraphics[width=\textwidth]{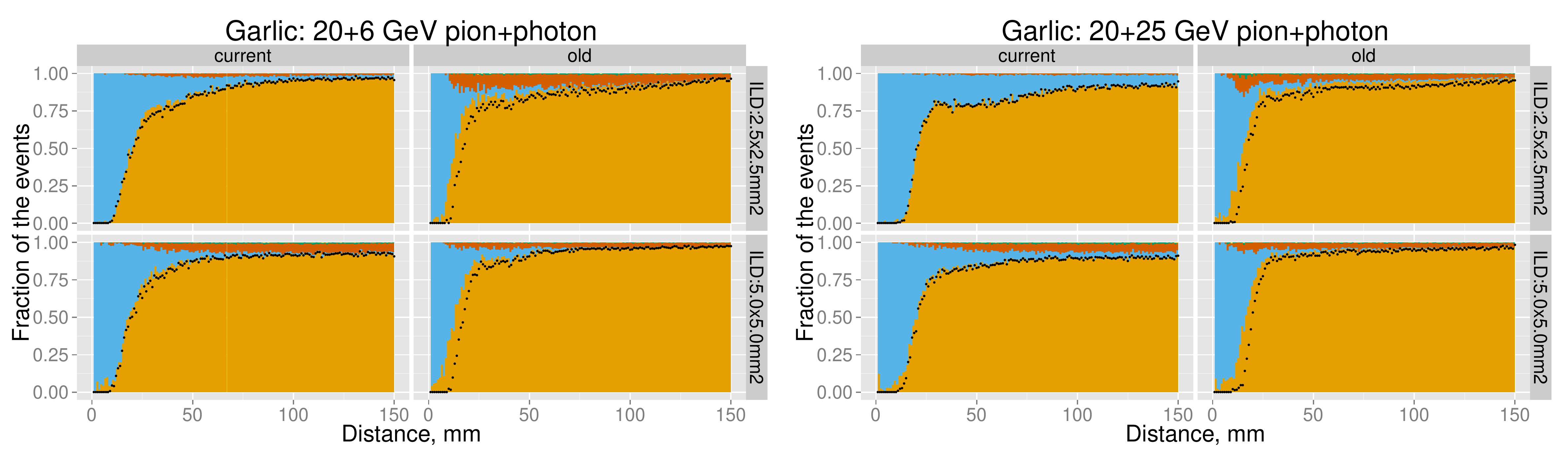}
 	\includegraphics[width=\textwidth]{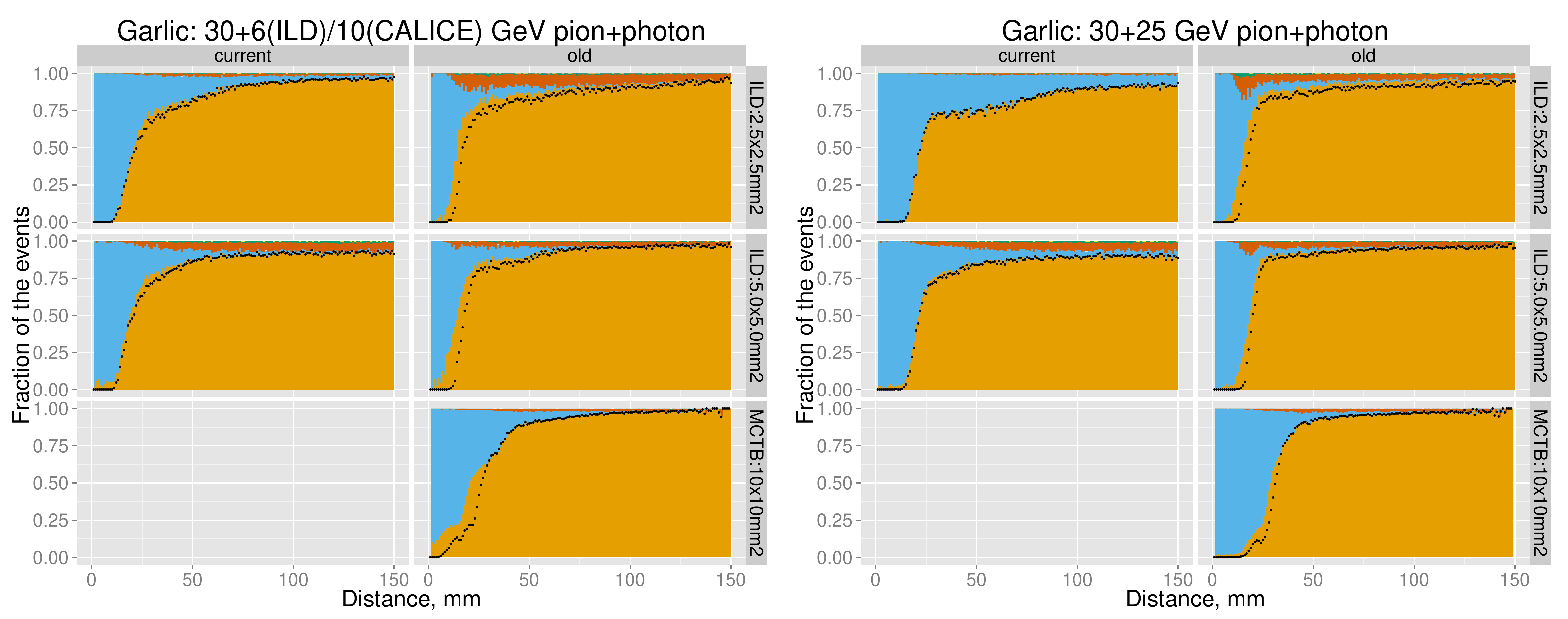}
 	\end{minipage}
 	\begin{minipage}{0.06\linewidth}
 	\includegraphics[align=c, width=\textwidth]{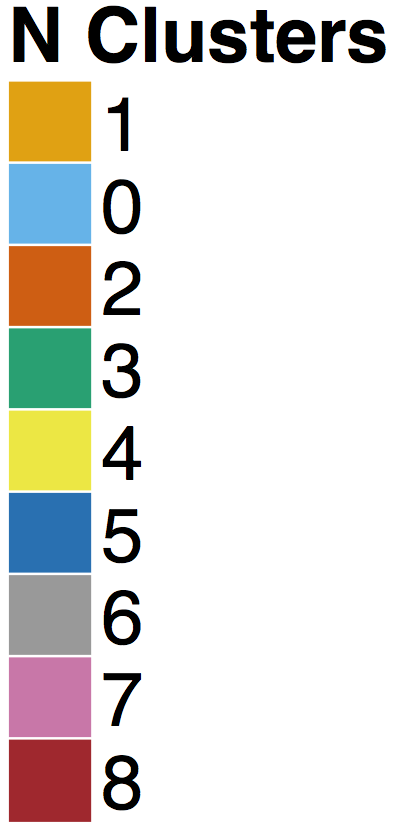}
 	\end{minipage}
 	\caption{Garlic (no HCAL) reconstruction of 20+6, 20+25, 30+6~(10) and 30+25~GeV $\pi^{+}-\gamma$ events: the colored bands show the fraction of events where 0, 1, 2,... photons are reconstructed, regardless of their energies and positions. Note, that Garlic reconstructs only ECAL hits and, in addition, hits around the pion track are vetoed and not considered by Garlic.}
 	\label {pionphotonGarlic}
\end{figure*}

Quantitatively all inefficiencies in $\gamma-\gamma$ and $\pi^{+}-\gamma$ reconstruction for three studied ECAL granularities ($2.5 \times 2.5$~mm$^2$, $5 \times 5$~mm$^2$ and $10 \times 10$~mm$^2$) and for all used versions of PFA programs are presented in Figs.~\ref{2photonsArbor}-\ref{pionphotonGarlic}. Colored bands denote the fraction of the events where 0, 1, 2 or more neutral clusters are reconstructed regardless of the PID, energy and barycenter position. All correctly reconstructed events in case of $\gamma - \gamma$ ($\pi^{+} - \gamma$) belong to the band where 2 (1) neutral clusters are reconstructed, this band is lowermost and is shown in orange. Black dots denote the separation efficiency shown earlier in Figs.~\ref{separationphotonTB}-\ref{separationhadronsILD}. Events, that belong to orange band, but above the efficiency curve, are not considered as correctly reconstructed due to at least one reason from the following: failure of $\gamma$ PID, reconstructed $\gamma$ energy (barycenter) are not within $\pm 20 \%$ ($\pm 5$~mm) of the initial reconstructed $\gamma$ values. These plots give a clearance of the main reconstruction problems that exist for each studied PFA program. Overall, it is seen that $\gamma - \gamma$ separation problems of Pandora v00-14, like over-split and merged photons (see, for example, Fig.~\ref{2photonsPandora} for 32+25~GeV performance), are resolved in v02-04. Performance of Garlic in case of nearby photons (photon close to hadron hits after track veto) is limited by the impossibility to consider a reconstructed cluster with the not perfect shape as photons. Arbor tends to over-split clusters in ECAL and in HCAL, this feature should be tuned in the future. Fig.~\ref{pionphotonPandora} for $\pi^{+}-\gamma$ events reconstructed with Pandora allow comparing efficiencies of the reclustering algorithm in case of $\pi^{+}-\gamma$ separation. For larger $\gamma$ energies algorithm better resolves $\pi^{+}-\gamma$ events, and the blue band corresponding to zero reconstructed photons is almost negligible in the case of 30+25~GeV (contrary to 30+6~GeV). The reclustering algorithm is not implemented in the used version of Arbor, and its performance can be additionally improved by this algorithm.

\section{Summary}

The difference between three programs reflect their different optimization goals: Pandora currently provides the best overall jet energy resolution, Garlic is a specialized tool for photon reconstruction in ECAL and Arbor attempt to reconstruct shower using a model of a tree. Performances of the algorithms were improved due to these studies.

\end{document}